# Emergence of unconventional magnetic order in strain-engineered $RuO_2/TiO_2$ superlattices

Seung Gyo Jeong[1,*], Seungjun Lee[2], Jin Young Oh[3], Bonnie Y.X. Lin[4], Anand Santhosh[1], James M. LeBeau[4], Alexander J. Grutter[5], Woo Seok Choi[3], Tony Low[2], Valeria Lauter[6,*], and Bharat Jalan[1,*]

[1]Department of Chemical Engineering and Materials Science, University of Minnesota–Twin Cities, Minneapolis, Minnesota 55455, USA

[2]Department of Electrical and Computer Engineering, University of Minnesota, Minneapolis, Minnesota 55455, USA

[3]Department of Physics, Sungkyunkwan University, Suwon 16419, Republic of Korea

[4]Department of Materials Science and Engineering, Massachusetts Institute of Technology, Cambridge, Massachusetts 02139, USA

[5]NIST Center for Neutron Research, National Institute of Standards and Technology, Gaithersburg, Maryland 20899, USA

[6]Neutron Scattering Division, Oak Ridge National Laboratory, Oak Ridge, Tennessee 37831, USA

[*]Corresponding authors: jeong397@umn.edu; lauterv@ornl.gov; bjalan@umn.edu

**Abstract**

The spin ordering in $RuO_2$ remains a highly debated topic, owing to its elusive nature, with reports ranging from a nonmagnetic ground state to signatures of unconventional magnetic order. Here we provide the first unambiguous, and direct evidence of unconventional magnetism in epitaxial, fully strained $RuO_2/TiO_2$ superlattices on $TiO_2$ (110) substrate grown by hybrid molecular beam epitaxy. Polarized neutron reflectometry reveals a finite magnetic moment localized within the compressively strained $RuO_2$ layers, consistent with predictions obtained from first-principles calculations. Complementary density functional theory and X-ray photoemission spectroscopy show that epitaxial strain drives the Ru $4d$ states toward the Fermi level, triggering a Stoner-type instability that stabilizes non-compensated magnetic order. These unique results reveal that $RuO_2$ exhibits unconventional magnetic states under epitaxial strain, which are not accessible in bulk and establish strain engineering as a powerful route to uncover and control magnetic phases in $RuO_2$ and related oxides.

**Introduction**

The magnetic state of $RuO_2$ has remained a subject of vigorous debate. Rutile $RuO_2$ has been proposed as a *d*-wave altermagnet[1,2], characterized by rotated spin densities, staggered spin polarizations, and a Néel vector aligned along [001] crystallographic direction. Early neutron and resonant X-ray diffractions reported a magnetic Bragg peak at the (100) reflection in bulk crystals and epitaxial films[3,4], suggesting a small, antiferromagnetically (AFM) ordered moment persisting above room temperature. Yet, later studies demonstrated that this reflection could arise from the intrinsic charge anisotropy of the rutile lattice[5,6], casting doubt on these earlier interpretations. Complementary muon spin rotation/relaxation (μSR) measurements have also produced conflicting results – initial experiments detected no magnetic moments in bulk crystals and sputtered epitaxial films[7,8], whereas more recent low-energy μSR with approximately 10 nm depth resolution revealed enhanced muon asymmetry near the surface in epitaxial films[9]. Angle-resolved photoemission spectroscopy (ARPES) has likewise yielded inconsistent observations across sample geometries and different synthesis techniques and sample geometries[10-15].

The diverging experimental reports highlight the importance of synthesis parameters such as sample geometry, dimensionality, defect chemistry, and epitaxial strain. While $RuO_2$ bulk is predominantly reported to be nonmagnetic[16-20], $RuO_2$ thin films show a significantly broader spectrum of magnetic responses[12,21-42]. These imply that the reduced dimensionality and/or interfacial constraints may substantially influence the structural, electronic, and magnetic ground states. Among these factors, epitaxial strain – owing to the lattice mismatch between $RuO_2$ films and $TiO_2$ (110) substrates – plays a particularly crucial role (Extended Data Fig. 1). Especially, compressive strain along [001] can shift the Ru 4*d* states toward the Fermi level, enhancing the Stoner instability[22,23], analogous to strain-driven magnetism in monolayer $SrRuO_3$[43]. Fully strained (110)-oriented $RuO_2$ films on $TiO_2$ (110) further break the two-fold spin-rotation symmetry generated by four-fold rotation combined with a half translation, giving rise to an uncompensated magnetic order, i.e., weak ferromagnetism (FM)[21-23].

Even small variations in strain state across differently grown samples can therefore yield qualitatively distinct magnetic responses, underscoring the need for a robust observation of magnetism in *fully strained* $RuO_2$ films on $TiO_2$ (110) substrates. Yet, a direct magnetization measurement of such films has remained elusive. Heterostructures incorporating a ferromagnetic layer have been explored to probe the magnetism of $RuO_2$ via spin torque or exchange bias measurements[25-39,41,44], such approaches are inherently affected by proximity effects between the ferromagnetic layer and $RuO_2$, making it difficult to isolate the magnetization response of $RuO_2$ itself. Moreover, experimental investigations on ultrathin, fully strained films are highly limited, likely because disorder and defects suppress the intrinsic metallicity and any emergent magnetic behavior. Hybrid molecular beam epitaxy (MBE) grown, fully strained $RuO_2/TiO_2$ (110) films have shown a robust metallicity and crystalline quality of the $RuO_2$ layer with several promising signatures of magnetism, including time-reversal symmetry breaking in optical second-harmonic generation[21], magneto-optical responses[21], anomalous Hall effects[23], and spin-split bands in spin-ARPES[12]. Notably, the observation of an anomalous Hall effect under magnetic fields applied along both out-of-plane [110] and in-plane [001] directions suggests the possible spin canting in this system[23]. However, these experiments do not directly quantify net magnetization; therefore, a direct measurement of the magnetic order is still lacking.

Here, we directly examine the uncompensated magnetic order in fully strained $RuO_2$ superlattices using polarized neutron reflectometry (PNR). PNR offers sub-nm depth resolution and isolates the layer-selective magnetic asymmetry characteristic of uncompensated order[45-52], separating magnetic scattering from structural contribution. To maximize magnetic contrast, we synthesized atomically well-defined $RuO_2/TiO_2$ superlattices on $TiO_2$ (110) substrates using hybrid MBE[12,21-23,49], achieving high signal-to-noise ratios on PNR experiment at the superlattice peaks while maintaining a fully strained state. By combining density functional theory (DFT) with X-ray photoemission spectroscopy (XPS), we reveal that epitaxial strain pushes the Ru 4*d* states toward the Fermi level, inducing Stoner instability. We observe clear signatures of uncompensated magnetic order in the strained $RuO_2$ layers – reproduced across multiple measurements and different samples – in agreement with DFT calculations. These results demonstrate that epitaxial strain stabilizes the uncompensated magnetic order in $RuO_2$ and that the magnetism is not confined to the surface only. These results unambiguously demonstrate how atomic-scale synthesis and strain engineering lead to unconventional magnetic states otherwise inaccessible in bulk form, providing a robust pathway for designing strain-engineered $RuO_2$ for next-generation spintronic and quantum technologies.

**Results and Discussion**

To obtain high signal-to-noise ratios at the superlattice peaks while maintaining epitaxial strain, we synthesized epitaxial $RuO_2$/$TiO_2$ (110) superlattice structures using hybrid MBE. We designed two superlattice variants with $RuO_2$ layer thicknesses of 2 and 3 nm – both below the ~ 4 nm critical thickness for strain relaxation[21] – while keeping the overall supercell thickness near 8.5 nm with five repetitions. Such design optimally places the primary superlattice reflection peak at a low out-of-plane wave vector ($q_{110}$), ensuring that multiple superlattice peaks fall within the accessible *q*-range for PNR. Streaky RHEED patterns after growth with pronounced Kikuchi lines along both the ⟨1 1̄ 0⟩ and ⟨001⟩ azimuths confirm excellent crystalline order and smooth surfaces (Extended Data Fig. 2).

High-resolution X-ray diffraction (XRD) *θ*–2*θ* scans and X-ray reflectivity (XRR) measurements verify the (110)-oriented superlattice structure (Figs. 1(a,b)). The $TiO_2$ (110)

Bragg peak appears at $q_{110} = 19.3$ nm$^{-1}$ with a distinct shoulder originating from the RuO$_2$ layers in both samples. Multiple superlattice satellites (SL$_{\pm n}$) confirm the well-defined RuO$_2$/TiO$_2$ periodicity, and the periodicities match the designed supercell dimensions. Pronounced and multiple Laue oscillations between Bragg reflections further indicate atomically sharp interfaces. Supercell thicknesses for two different samples extracted from XRD (8.6 and 8.5 nm, respectively) agree with values obtained from PNR fitting (discussed below). Based on PNR and XRR, we assign the layer sequences as [1.8 nm RuO$_2$|6.9 nm TiO$_2$]$_5$ and [2.8 nm RuO$_2$|5.6 nm TiO$_2$]$_5$ for the two samples, respectively. In the XRR scans, the multiple reflectivity oscillations of the superlattice persist up to $q_{110} = 5.5$ nm$^{-1}$, demonstrating the atomically smooth surface and low interface roughness of both superlattices. A representative atomic force microscopy (AFM) for [1.8 nm RuO$_2$|6.9 nm TiO$_2$]$_5$ sample further confirms atomically smooth surfaces, with a root mean square (RMS) roughness ($S_q$) of ~0.2 nm (Inset of Fig. 1b).

XRD reciprocal space maps (RSMs) around the TiO$_2$ (332) and (310) reflections confirm that both in-plane crystal orientations ([001] and [1$\bar{1}$0]) remain fully strained (Fig. 1c), emerging a nontrivial magnetic state in RuO$_2$ layers (Extended Data Fig. 1). For RuO$_2$/TiO$_2$ (110), the lattice mismatches between RuO$_2$ and TiO$_2$ are −4.7% along the [001] and +2.3% along the [1$\bar{1}$0]. In single-layer RuO$_2$ films, the RuO$_2$ Bragg reflection peak largely overlaps with that of the TiO$_2$ substrate, making it difficult to determine the strain state using laboratory-based XRD. However, in the superlattice geometry, the appearance of well-defined superlattice satellite peaks enables direct validation of identical in-plane wave vectors between superlattice and substrate, i.e., a fully strained state of RuO$_2$ layers.

Our DFT calculations indicate that the fully strained state shifts the Ru 4$d$ valence states closer to the Fermi level, thereby promoting the stabilization of a magnetic ground state following the Stoner model (Extended Data Fig. 1b). This strain-induced band shift is consistent with our XPS results on $RuO_2$/$TiO_2$ (110) films revealing that the Ru 4*d* valence-band peak systematically moves toward the Fermi level with decreasing thickness (increasing strain) (Extended Data Fig. 1d). The DFT calculations further reveal an uncompensated magnetic order in fully strained epitaxial $RuO_2$/$TiO_2$ (110) system (Extended Data Fig. 1c).

The PNR measurements directly confirm the presence of the net magnetization in the fully strained $RuO_2$ layers of the superlattices (Fig. 2). Figure 2a shows the PNR spectra of the [2.8 nm $RuO_2$|5.6 nm $TiO_2$]$_5$ superlattice measured at 3.7 K under a 4.8 T magnetic field applied along the [001] direction. $R^+$ and $R^-$ denote the PNR spectrum measured with the neutron spin parallel and antiparallel to the applied magnetic field, respectively. To suppress potential domain formation and enhance magnetic contrast, the sample was field-cooled (FC) from room temperature in a 4.8 T field along [001], and measurements were performed at a temperature ($T$) < 4 K, which is lower than the theoretically predicted magnetic anisotropy energy (~13 K)[23]. The insulating $TiO_2$ layers of the superlattices are thick enough that any direct exchange coupling between $RuO_2$ layers should be strongly suppressed[53], so the observed signal is interpreted as a layer-confined $RuO_2$ magnetization aligned by the applied field. Similar to XRR, PNR exhibits well-defined Kiessig fringes and superlattice peaks at $q_{110}$ = 0.75 and 1.48 nm$^{-1}$, reflecting excellent crystallinity and supercell thickness. Magnified views near the Bragg reflections (inset of Fig. 2a) reveal a clear separation between $R^+$ and $R^-$ beyond the experimental uncertainties, clearly demonstrating the existence of uncompensated magnetic order in the fully strained $RuO_2$/$TiO_2$ superlattices. To further confirm reproducibility, we measured two different superlattice samples – [1.8 nm $RuO_2$|6.9 nm $TiO_2$]$_5$ and [2.8 nm $RuO_2$|5.6 nm $TiO_2$]$_5$ – within a single beamtime session (Extended Data Fig. 3), and performed additional measurements on the [1.8 nm $RuO_2$|6.9 nm

TiO$_2$]$_5$ sample during a separate beamtime (Extended Data Fig. 4). All three independent datasets consistently show robust splitting between R$^+$ and R$^-$ at the superlattice peaks, firmly establishing the presence of uncompensated magnetic order. In contrast, no observable magnetic splitting is seen near the superlattice peaks when the measurement is performed along the [1$\bar{1}$0] orientation (Extended Data Fig. 5), suggesting that the uncompensated magnetic component lies along the [001] direction. We also note that the PNR method is a three-dimensional depth-resolved vector magnetometry sensitive to the in-plane magnetization vector[54,55]. To separate the nuclear from the magnetic scattering, the data is presented as the spin-asymmetry (SA) ratio SA = (R$^+$($q_{110}$) − R$^-$($q_{110}$))/(R$^+$($q_{110}$) + R$^-$($q_{110}$)), as depicted in Fig. 2b, and highlights the splitting near the superlattice reflections, where the reflectivity signal is maximal, and the statistical uncertainties are minimal. The SA near the critical edge at $q_{110}$ = 0.16 nm$^{-1}$ also exhibits a small but robust spin splitting exceeding statistical uncertainties, further supporting the presence of a net magnetization in the superlattices. In contrast, the small SA signal away from the superlattice peaks is below the measurement uncertainty, precluding a detailed spin-dependent analysis in these regions.

To quantify the magnetic signal, we fitted the PNR spectra using a minimal superlattice model (see Extended Data Fig. 6 and Extended Data Table 1 for a more detailed model analysis), intentionally avoiding overly complex parameterizations. The resulting fits (solid lines in Figs. 2a and 2b) show good agreement with the experimental data, particularly in the vicinity of the superlattice reflections where the magnetic contrast is most pronounced. The $\chi^2$ value of 2.50 further indicates that the deviations between the experimental data and the simulations remain within statistical expectations. Figures 2c and 2d present the nuclear and magnetic scattering length densities (SLDs) extracted using Ref1D, revealing an atomically well-defined periodic structure and a finite uncompensated magnetic moment confined to each RuO$_2$ layer. The nuclear SLD of the RuO$_2$ layers is (5.90 ± 0.01) × 10$^{-4}$ nm$^{-2}$, in excellent agreement with the theoretical bulk RuO$_2$ density of $5.88 \times 10^{-4}$ nm$^{-2}$. The magnetic SLD of the RuO$_2$ layers is $2.19 \times 10^{-6}$ nm$^{-2}$, corresponding to the averaged magnetization of $(0.0219 \pm 0.0039)$ $\mu_B$ f.u.$^{-1}$. To test alternative scenarios, we simulated several possible magnetic configurations, including conventional FM in RuO$_2$ (Extended Data Fig. 7a) and magnetism residing in the TiO$_2$ layers (Extended Data Fig. 7b). These models produce signatures that clearly differ from the experimental data; in particular, magnetic TiO$_2$ layer would yield an opposite sign of spin splitting at the superlattice peaks (Extended Data Fig. 7b), which is incompatible with our observations. PNR fits obtained for another sample [1.8 nm RuO$_2$|6.9 nm TiO$_2$]$_5$ superlattice also confirm the presence of uncompensated magnetic order in each RuO$_2$ layer (Extended Data Fig. 6b). While the spin splitting at the second superlattice peak of [1.8 nm RuO$_2$|6.9 nm TiO$_2$]$_5$ superlattice consistently indicates net magnetization in the RuO$_2$ layers, the difference in spin splitting between the first and second superlattice peaks may suggest an enhanced contribution from interfacial regions (Extended Data Fig. 8), which we attribute to the reduced RuO$_2$ thickness.

DFT calculations on RuO$_2$/TiO$_2$ superlattices reveal net magnetization values that agree closely with the PNR results (Fig. 3). Figures 3a and 3b summarize the key quantities extracted from experiment and theory, respectively. To quantitatively assess uncertainties and establish confidence intervals, we performed the DREAM Markov-chain Monte Carlo (MCMC) algorithm implemented in the bumps Python package[50,51,56] (Extended Data Table 1). The net magnetization values obtained from PNR experiments are $(0.0219 \pm 0.0039)$ and $(0.0111 \pm 0.0080)$ $\mu_B$ f.u.$^{-1}$ for the [2.8 nm RuO$_2$|5.6 nm TiO$_2$]$_5$ and [1.8 nm RuO$_2$|6.9 nm TiO$_2$]$_5$ superlattices, respectively (Fig. 3a). These results support the presence of an uncompensated magnetization along the [001] direction of the fully strained RuO$_2$ layers of the superlattices, beyond the bounds of fitting uncertainty. It is worth comparing our results with the expected magnetic field-induced paramagnetic moments. Assuming a molar susceptibility of $RuO_2$ of $1.5 \times 10^{-4}$ emu mol$^{-1}$ Oe$^{-1}$ at 4 K[3], the moment induced by a 4.8 T magnetic field is only ~0.0013 $\mu_B$ f.u.$^{-1}$. Thus, the net moment we observe along the [001] direction cannot be explained by paramagnetism alone. For the DFT calculations, we adopted a fully strained [2 nm $RuO_2$|2 nm $TiO_2$]$_5$ superlattice without a Hubbard $U$ correction ($U = 0$). The calculated total magnetization of the uncompensated AFM (weak FM) state is 0.0288 $\mu_B$ f.u.$^{-1}$, in good quantitative agreement with the PNR values, whereas the FM state yields a much larger moment of 0.2820 $\mu_B$ f.u.$^{-1}$ (left panel of Fig. 3b).

The relative total energies of the AFM (weak FM), FM, and nonmagnetic (NM) configurations (right panel of Fig. 3b) establish the AFM state as the magnetic ground state (the lowest energy), consistent with earlier $RuO_2$ studies[12,21-24]. We further note that the experimentally extracted magnetization values also rule out the FM state (Extended Data Fig. 6b). These results align with recent observations in fully strained $RuO_2$/$TiO_2$ (110) films, including spin-split bands in spin-ARPES[12], time-reversal symmetry breaking in optical second-harmonic generation[21], and anomalous Hall effect[23]. Group theory analysis further indicates that the uncompensated magnetic phase hosts a coexistence of weak FM and altermagnetism[21,57], while the theoretical calculations reveal a non-collinear spin texture carrying both octupolar (AFM/altermagnetic) and non-zero net magnetic (weak FM) moments[21]. While conventional PNR is insensitive to fully compensated magnetic states - where opposing magnetic moments cancel each other out perfectly (e.g., in a bulk antiferromagnet)[58]— it possesses high sensitivity to weak uncompensated moments – making it ideally suited for detecting and quantifying magnetic states stabilized in our strained $RuO_2$ layers within the superlattice.

Magnetic anisotropy energy calculations indicate that the magnetic easy axis lies along the in-plane [001] direction, rather than the out-of-plane [110] direction (Fig. 3c). The atom-resolved spin-density distribution (Fig. 3d) shows that the magnetic moments originate predominantly from Ru atoms, consistent with the layer-resolved magnetic SLD obtained from PNR analysis (Fig. 2d). Notably, the projected magnetic moment profile in Fig. 3e reveals a pronounced interfacial enhancement of the magnetic order near the $RuO_2$/$TiO_2$ boundaries, which might be related to the interfacial magnetic features observed in the [1.8 nm $RuO_2$|6.9 nm $TiO_2$]$_5$ superlattices (Extended Data Fig. 8). This feature is reminiscent of the theoretically predicted interfacial magnetic behavior at the boundary between the dielectric $SrTiO_3$ and the spin-polarized metal $SrRuO_3$[59].

In summary, our results directly observe the magnetic order in $RuO_2$/$TiO_2$ superlattices via PNR, combining with first-principles calculations. We have established (1) a strain-stabilized magnetic state driven by the epitaxial strain, and (2) the emergence of uncompensated magnetic order originating from strain-induced crystal distortions. Our approach targets the uncompensated magnetic component using PNR, providing the first unambiguous and direct evidence of net magnetization and addressing the key aspect of the ongoing debate regarding the magnetic state of strained $RuO_2$. These results establish epitaxial strain as a powerful and versatile control knob for engineering, accessing, and controlling magnetic phases in correlated oxides.

**Methods**

**Polarized Neutron Reflectometry**

PNR experiments were performed on the Magnetism Reflectometer MAGREF at the Spallation Neutron Source at Oak Ridge National Laboratory[60-63], using neutrons with wavelengths λ in the range of 0.2–0.8 nm and a high degree of polarization of 98.5–99%. The measurements were performed in a closed cycle refregirator equipped with a 5 T cryomagnet (Cryomagnetics). Using the time-of-flight method, a beam of collimated polychromatic polarized neutrons with a wavelength band of δλ impinged onto a film at a grazing angle θ, interacting with atomic nuclei and the spins of unpaired electrons. The reflected intensity R+ and R− was measured as a function of momentum transfer, $q_{110} = 4\pi \sin(\theta)/\lambda$, with the neutron spin parallel (+) or antiparallel (−) to the applied field, respectively. To separate nuclear and magnetic scattering, the spin asymmetry ratio $SA = (R^+(q_{110}) - R^-(q_{110}))/(R^+(q_{110}) + R^-(q_{110}))$ was calculated, where SA = 0 corresponds to a zero magnetic moment in the system. Being electrically neutral, spin-polarized neutrons penetrate the entire multilayer structures and probe the magnetic and structural composition of the film and buried interfaces all the way down to the substrate.

**Superlattice Growth via Hybrid MBE**

$RuO_2/TiO_2$ (110) superlattices were synthesized on $TiO_2$ (110) single crystal substrates (Crystec) using a hybrid molecular beam epitaxy (MBE; Scienta Omicron). Substrates were cleaned sequentially with acetone, methanol, and isopropanol, followed by a 2-hour bake at 200 °C in the load-lock chamber. Before growth, substrates were annealed for 20 minutes in an oxygen plasma at 300 °C to ensure a clean surface. The thin $TiO_2$ buffer layer was incorporated to avoid substrate-induced variations and ensure consistent growth across different samples[23]. Each $RuO_2/TiO_2$ superlattice period was fabricated by alternating the cell shutter control of $RuO_2$ and $TiO_2$ layers. $RuO_2$ layers were grown using thermally evaporated $Ru(acac)_3$ delivered from a low-temperature effusion cell (MBE Komponenten) operated at 170–180 °C. $TiO_2$ layers were grown by using titanium tetraisopropoxide (TTIP, 99.999%, Sigma-Aldrich) injected through a line-of-sight gas injector (E-Science Inc.) controlled via a customized linear-leak valve system with a Baratron capacitance manometer (MKS

Instruments Inc.). All growth steps were performed under radio-frequency oxygen plasma at a pressure of 5 × 10⁻⁶ Torr. To suppress oxygen vacancy formation, the samples were cooled to 120 °C in oxygen plasma after completing the superlattice growth. Real-time growth evolution of surfaces was monitored through reflection high-energy electron diffraction (RHEED) before, during, and after growth.

**Structural Characterization**

Structural properties of the $RuO_2/TiO_2$ superlattices–including crystallinity, superlattice periodicity, thickness, interface sharpness, and strain state–were characterized using high-resolution X-ray diffraction (XRD; Rigaku SmartLab XE) $\theta$–$2\theta$ scans and X-ray reflectivity (XRR) were used to determine the superlattice period and interfacial roughness, while reciprocal space maps (RSMs) around asymmetric reflections were used to quantify the in-plane and out-of-plane strain states. Surface morphology was examined via atomic force microscopy (AFM, Bruker Nanoscope V Multimode 8) operated in peak-force tapping mode.

**Density Functional Theory**

We performed first-principles calculations based on density functional theory (DFT) as implemented in the Vienna *ab-initio* simulation package (VASP)[64,65]. We employed the projector augmented wave (PAW) method[66,67] and the Perdew-Burke-Ernzerhof (PBE) potential[68]. The kinetic energy cutoff of electronic wavefunctions was set to 500 eV. To describe (110)-oriented $RuO_2$, we aligned three lattice vectors of $RuO_2$ along [001], [$\bar{1}$10], and [110] directions, respectively. The theoretical equilibrium lattice constants of bulk $RuO_2$ and $TiO_2$ were calculated to be $a_{[001]}^{RuO2} = 3.119$ Å, and $a_{[\bar{1}10]}^{RuO2} = a_{[110]} = 6.394$ Å, and $a_{[001]}^{TiO2} = 2.970$ Å, and $a_{[\bar{1}10]}^{TiO2} = a_{[110]} = 6.588$ Å, respectively. For the $RuO_2/TiO_2$ superlattice structure, we constructed a supercell structure including six atomic sublayers of $RuO_2$ and six atomic sublayers of $TiO_2$, respectively. The in-plane lattice constants of the supercell are fixed to those of bulk TiO$_2$, while the out-of-plane lattice constants are fully relaxed, yielding a value of 39.5 Å. The corresponding Brillouin zone of the supercell was sampled using an 8 × 4 × 1 k-grid mesh. Spin–orbit coupling was included only in the calculation of the magnetic anisotropy energy.

**X-ray Photoelectron Spectroscopy**

Film composition was analyzed using X-ray photoelectron spectroscopy (XPS) with an Al K$\alpha$ source (1486.6 eV) and settings including an energy step size of 0.05 eV and a pass energy of 55 eV. A flood gun was adopted to prevent photoemission-induced surface charge effects. All spectra were calibrated using the binding energy of C–C bonding (284.8 eV) from the adventitious carbon on the surface.

**Data availability**

The data that support the findings of this study are available within the Article and its Supplementary Information. Other relevant data are available from the corresponding authors upon reasonable request.

**Code availability**

The PNR fitting codes used in this work are available at [https://github.com/reflectometry/refl1d] and [https://aglavic.github.io/genx/].

**Figures**

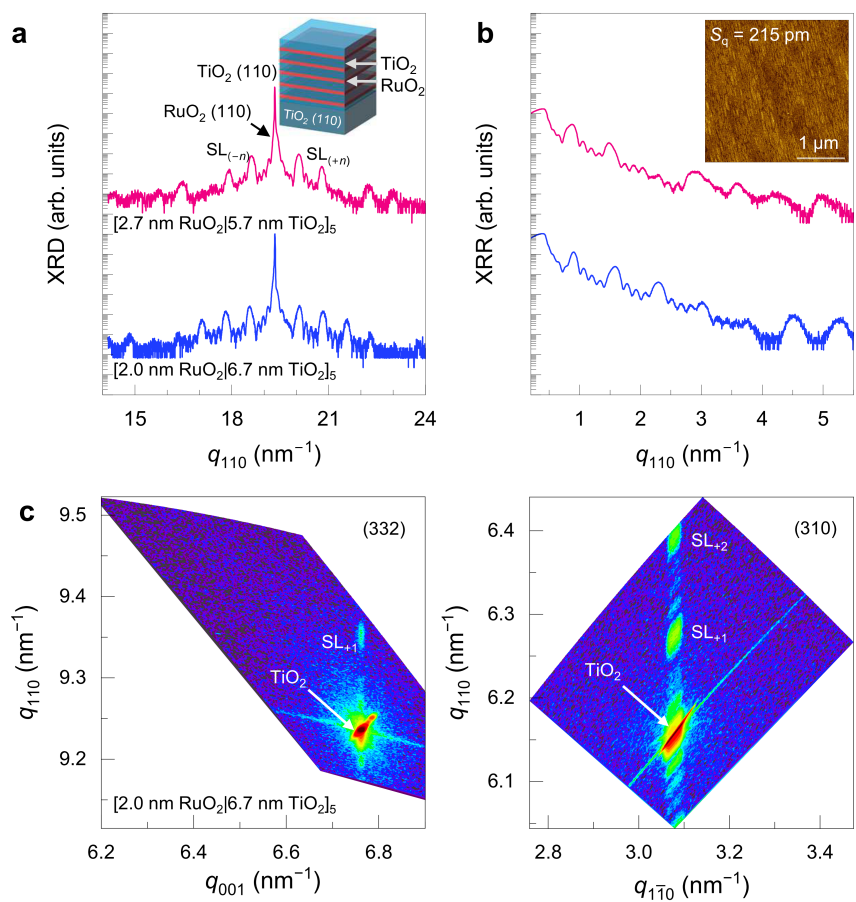

**Fig. 1. Atomically well-defined and fully strained RuO$_2$/TiO$_2$ (110) superlattices grown by hybrid molecular beam epitaxy. a,b,** High-resolution (**a**) XRD $\theta$–$2\theta$ and (**b**) XRR scans of the RuO$_2$/TiO$_2$ (110) superlattice thin films. Multiple superlattice satellite peaks (SL$_{\pm n}$) with Kiessig fringes in $\theta$–$2\theta$ scans and Laue oscillations in XRR scans indicate the intended periodicity and atomically sharp interfaces. The XRD $\theta$–$2\theta$ curves are vertically offset by $10^7$ and the XRR curves by $10^6$ for clarity. The inset of (**b**) shows an AFM image revealing an atomically smooth surface with a root-mean-square roughness of $S_q$ = 215 pm. **c,** XRD RSM scans around the (332) and (310) TiO$_2$ reflections show identical in-plane reciprocal space wave vectors between superlattice and substrate, indicating a fully strained state.

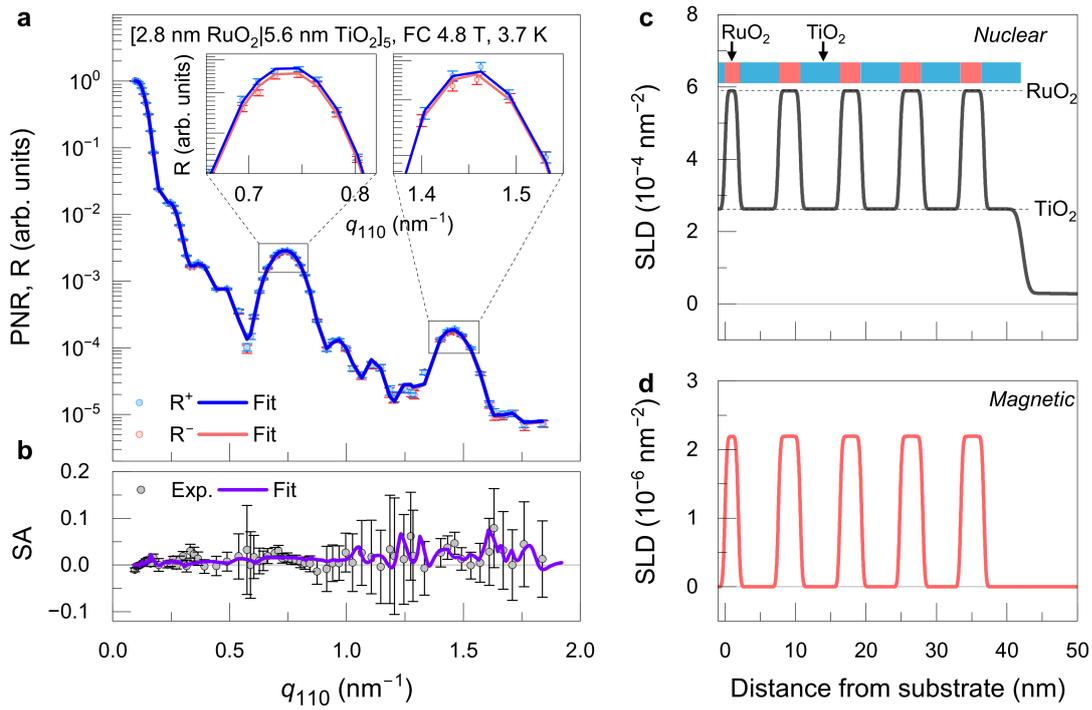

**Fig. 2. Uncompensated magnetic order in fully strained RuO$_2$/TiO$_2$ superlattices revealed by polarized neutron reflectometry. a,** Spin-resolved PNR spectra (R$^+$ and R$^-$) of the RuO$_2$/TiO$_2$ superlattice, together with the corresponding fits. Measurements were performed at 3.7 K under a 4.8 T magnetic field applied along the [001] direction. Insets show magnified views of the PNR spectra near the superlattice Bragg peaks, where the splitting between R$^+$ and R$^-$ becomes evident. **b,** Spin asymmetry, SA, along with the model fits. **c,d,** Depth profiles of the (**c**) nuclear and (**d**) magnetic scattering length densities (SLDs) used for reflectivity modeling. Horizontal dashed lines indicate the maximum SLD levels and serve as guides to the eye. The inset in (**c**) provides a schematic illustration of the superlattice stack.

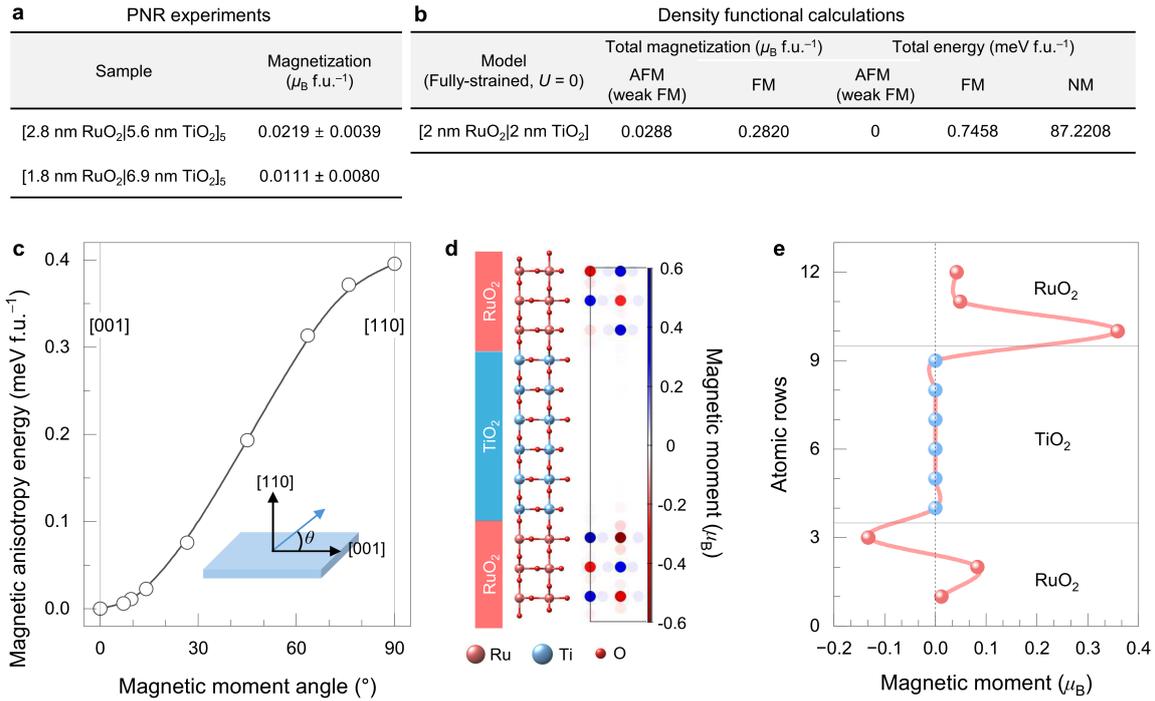

**Fig. 3. Experimental magnetization and theoretical description of uncompensated magnetic order in strained RuO$_2$/TiO$_2$ superlattices. a**, the magnetization values extracted from PNR measurements for two RuO$_2$/TiO$_2$ superlattice samples, demonstrating the presence of weak yet finite magnetic moments in the fully strained RuO$_2$ layers. **b**, DFT calculated relative total energies of fully strained RuO$_2$/TiO$_2$ superlattices ($U = 0$) for uncompensated antiferromagnetic (AFM; weak ferromagnetic), ferromagnetic (FM), and nonmagnetic (NM) states. The AFM (weak FM) configuration is energetically favored. **c**, Magnetic anisotropy energy of the AFM (weak FM) state as a function of the spin-axis orientation between the [001] and [110] crystallographic directions, identifying [001] as the magnetic easy axis. **d**, Atom-resolved spin-density distribution for the RuO$_2$/TiO$_2$/RuO$_2$ heterostructure (schematically shown on the left), highlighting Ru-derived magnetic moments. **e**, Projected magnetic moment profile plotted across the atomic rows of the heterostructure in (**d**).

**Acknowledgements**

Film synthesis and structural characterizations (S.G.J. and B.J.) were supported by the U.S. Department of Energy through grant Nos. DE-SC0020211, and (partly) DE-SC0024710. Transport, and ellipsometry (at UMN) were supported by the Air Force Office of Scientific Research (AFOSR) through Grant No. FA9550-21-1-0025 and FA9550-24-1-0169. Film growth was performed using instrumentation funded by AFOSR DURIP awards FA9550-18-1-0294 and FA9550-23-1-0085. S.G.J, B.J, S.L, and T.L also benefited from the Air Force Office of Scientific Research Multi University Research Initiative (AFOSR MURI, Award No. FA9550-25-1-0262). Parts of this work were carried out at the Characterization Facility, University of Minnesota, which receives partial support from the NSF through the MRSEC program under Award No. DMR-2011401. This research used resources at the Spallation Neutron Source, a Department of Energy Office of Science User Facility operated by the Oak Ridge National Laboratory. This work was supported by the National Research Foundation of Korea funded by the Korean government (2021R1A2C2011340, RS-2023-00220471, and RS-2023-00281671). Certain commercial equipment, instruments, software, or materials are identified in this paper in order to specify the experimental procedure adequately. Such identifications are not intended to imply recommendation or endorsement by NIST, nor it is intended to imply that the materials or equipment identified are necessarily the best available for the purpose. Unless otherwise noted, work at NIST was funded solely by the United States government. This manuscript has been authored by UT-Battelle, LLC under Contract No. DE-AC05-00OR22725 with the U.S. Department of Energy. The United States Government retains and the publisher, by accepting the article for publication, acknowledges that the United States Government retains a non-exclusive, paid-up, irrevocable, world-wide license to publish or reproduce the published form of this manuscript, or allow others to do so, for United States Government purposes. The Department of Energy will provide public access to these results of federally sponsored research in accordance with the DOE Public Access Plan [https://www.energy.gov/downloads/doe-public-access-plan]. B.L. and J.M.L. were supported by the AFOSR through grant No. FA9550-20-1-0066.

**Contributions**

S.G.J. and B.J. conceived the idea and established proof of concept. S.G.J. and A.S. grew and characterized them using RHEED, XRD and AFM. S.L. and T.L. performed DFT calculations. S.G.J., J.Y.O., W.S.C., and V.L. performed PNR measurement and S.G.J., A.J.G., V.L. and B.J. analyzed PNR results. B.J. directed and organized the different aspects of the project. S.G.J. and B.J. wrote the manuscript. All authors contributed to the discussion and preparation of the manuscript.

**Corresponding authors**

Correspondence to Seung Gyo Jeong or Valeria Lauter or Bharat Jalan.

**Competing interests**

The authors declare no competing interests.